# ZPAN: An Organic Nonlinear Optical Crystal for High Intensity THz Generation

**Matthew J. Lutz*, (Enoch) Sin-Hang Ho*, Zachary B. Zaccardi[1], Paige Petersen[1], Elisha Jones[1], Stacey J. Smith[1], David J. Michaelis,[a)] Jeremy A. Johnson,[a)]**

*Contributed equally

Department of Chemistry and Biochemistry, Brigham Young University, Provo, UT 84602, USA

[a)] **Authors to whom correspondence should be addressed**: djm69@byu.edu, jjohnson@byu.edu

We report an optimized synthesis and crystal growth protocol as well as structural, optical, and terahertz (THz) generation characteristics of the organic nonlinear optical crystal ZPAN ((Z)-1-(((4-Phenylamino)phenylamino)methylene)naphthalen-2(1H)-one). Noncentrosymmetric packing of ZPAN is most reliably achieved by slow evaporation from an acetone-based solvent system, producing long rectangular prisms (7 cm × 1 cm × 0.2 cm) with (010) as the main face and [001] as both the polar axis and the direction of elongated crystal growth. [001] is thus the most effective pump polarization direction for generating THz light via optical rectification. The THz generation characteristics of the developed ZPAN crystals are determined at different near-infrared irradiation wavelengths, crystal thicknesses, and pump powers, revealing that ZPAN can generate a peak-to-peak electric field of near 1 MV/cm with a smooth spectrum from 0.5-3.4 THz. Calculated second order nonlinear optical coefficients indicate that both the (100) and (010) faces are theoretically capable of THz generation, though (010) is consistently the dominant growth face. The large size of this face (7 cm × 1 cm) makes it possible to scale laser power with aperture size, for use in extremely high-power laser systems.

## INTRODUCTION

In recent years, the use of terahertz radiation in optical and electronic applications has led to advancements in many fields including imaging, computing, security, and quality control [1-3]. The increase in applications requiring high intensity terahertz (THz) light has spurred the development of new materials capable of THz generation. One of the most efficient methods of THz generation is through optical rectification of near-infrared (NIR) laser light in nonlinear optical (NLO) organic crystals, which produce broad spectrum and high intensity THz light [4-6]. Each organic NLO material produces THz light with unique spectral features and intensities that can be useful in different applications. Some of the more powerful and practical organic THz generators reported to date include PNPA [7], DAST and DSTMS [8], OH1 [9], BNA [10], F-BNA [11], and MNA [12].

Traditional approaches to discovering new organic NLO materials are laborious, requiring the optimization of multiple material properties [5]. For example, organic THz-generating materials must have high molecular hyperpolarizabilities but must also have noncentrosymmetric alignment of molecules in the crystal structure. In our recent work, we demonstrated that data-mining the Cambridge Structural Database (CSD) for organic materials with high molecular hyperpolarizability and noncentrosymmetric arrangements is an efficient approach to discovering new THz generation materials [6]. One of the materials we identified is (Z)-1-(((4-Phenylamino)phenylamino)methylene)naphthalen-2(1H)-one, or ZPAN. The original structural report of ZPAN computationally predicted it would have high hyperpolarizability and thus strong potential as a new NLO material [13]. In this work, we experimentally investigate ZPAN as an efficient THz generation material and provide an optimized synthesis and crystal growth protocol, as well as a more complete report on its structure and THz spectra than our prior work [6]. One significant finding is that with the optimized synthetic conditions, ZPAN crystals can be grown with high internal quality to much larger sizes (up to 7 cm × 1 cm) than other THz generators, making it particularly attractive for use with very high-pulse energy laser systems in the 1-3 THz range.

## EXPERIMENTAL

### Synthesis and Crystal Growth

ZPAN was synthesized on large scale as follows. 4-Aminodiphenylamine (14.84 g) and 2-hydroxy-1-naphthaldehyde (13.97 g) were dissolved in ethanol until no solids remained (~300 ml). The solution was then refluxed at 85 °C for 2 hours. The target compound began precipitating during the reaction. The reaction mixture was cooled to room temperature and the solvent removed under reduced pressure on a rotary evaporator. The resulting red solids were dissolved in a minimal amount of chloroform (~300-400 ml) with heating and then placed in a freezer to recrystallize. The solids were collected via filtration. Recrystallization a second time from chloroform was used to obtain high purity products. The solids were then dried under vacuum to give a 94% yield. $^{1}$H NMR (500 MHz, $CDCl_3$), δ 15.76 (s, 1H), 9.32 (s, 1H), 8.10 (d, $J$ = 8.5 Hz, 1H), 7.78 (d, $J$ = 9.2 Hz, 1H), 7.71 (d, $J$ = 7.9 Hz, 1H), 7.52 (t, $J$ = 7.9 Hz, 1H), 7.34-7.29 (m, 5H), 7.14-7.09 (m, 5H), 6.98 (t, $J$ = 7.4 Hz, 1H), 5.84 (s, 1H).

While previous studies suggested that single crystals of ZPAN grow via slow evaporation from ethanol [13], crystals grown via this method in the present work resulted in centrosymmetric molecular packing. We found that slow evaporation of the purified ZPAN from acetone solutions gives a consistent growth of crystals with noncentrosymmetric packing in space group $Pna2_1$. We also found that using acetone and water as a mixed solvent system enables the growth of larger crystals. Noncentrosymmetric ZPAN crystals in this study were obtained by slow evaporation from a mixed solvent (acetone and 5 vol% water) in a crystallization dish at an evaporation rate of about 2 ml per day**. Fig. 1(a)** shows the long, rectangular ZPAN crystals that were extracted after several days of growth, with raw crystals longer than 7 cm, wider than 1 cm, and ranging from 0.5 to 2 mm thick. The crystals can be polished to optimal thicknesses as illustrated in **Fig. 1(b)**. The ease of growing large, high-quality crystals makes ZPAN an excellent candidate for terahertz generation with exceedingly large pulse energies. **Fig. 1(c)** shows the molecular structure for the ZPAN molecule.

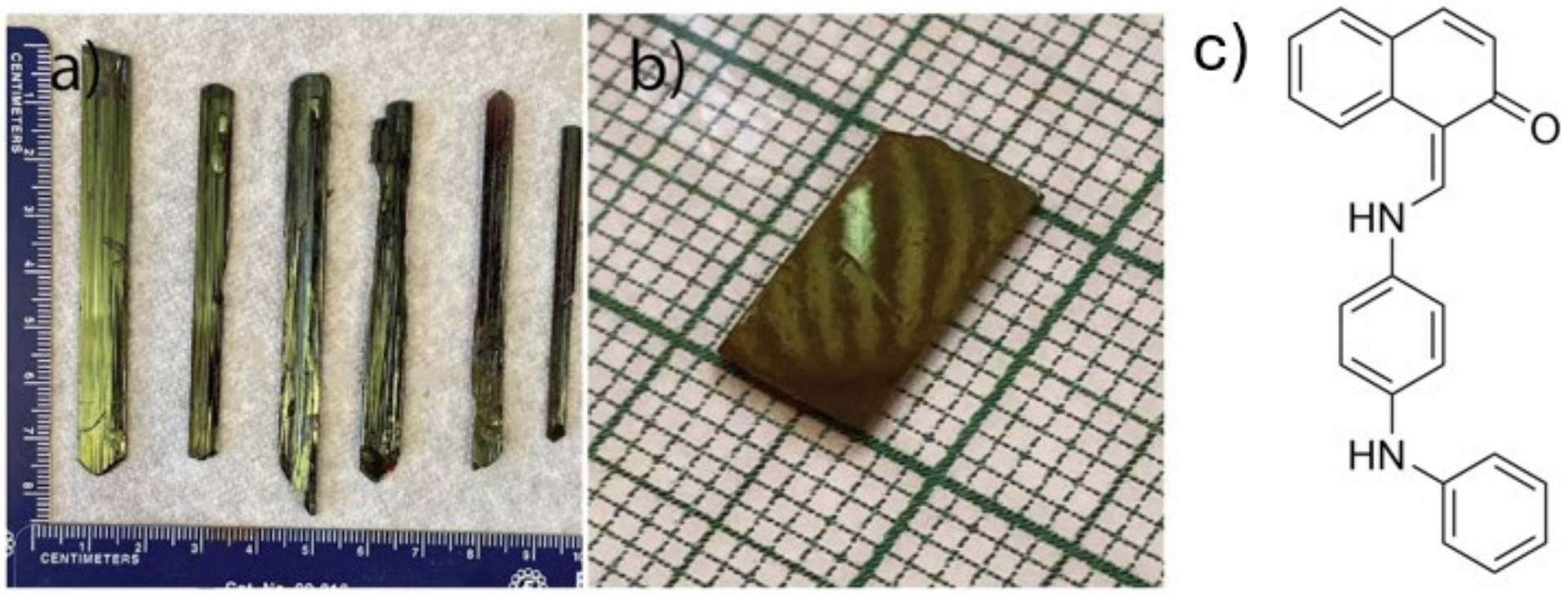

Figure 1: a) ZPAN crystals after initial growth, with the largest crystal being 7.5 cm x 1 cm x 0.1 cm. b) a 1 cm x ~0.5 cm x ~0.1 cm ZPAN crystal after processing by cutting into smaller sections, cleaving, and polishing. c) The molecular structure of a ZPAN molecule.

**Crystal Structure and Face Determination**

The structures of the ZPAN crystals were determined through single-crystal X-ray diffraction (SC-XRD). Red ZPAN crystals were cut to $0.098 \times 0.244 \times 0.692$ mm$^3$ for the noncentrosymmetric and $0.15 \times 0.15 \times 0.05$ mm$^3$ centrosymmetric materials. Each was mounted on a polyimide loop and centered on the goniometer under a stream of cold $N_2$ gas. Low-temperature (100 K) X-ray diffraction data were collected via φ- and ω- scans using a Bruker D8 Venture equipped with a kappa goniometer, dual microfocus X-ray sources, a Bruker Photon-III detector, and an Oxford 800+ low temperature device. Cu Kα radiation (λ = 1.54178 Å) was used for each crystal. Reflections were processed using the Bruker APEX6 suite applying absorption corrections via a multi-scan method [14]. The crystal structures were solved in space groups $Pna2_1$ and *Pbcn*, respectively, using SHELXT [15] and refined against $F^2$ by full-matrix least squares with SHELXL-2014 [16]. Anomalous scattering was used to calculate a Flack parameter (-0.051) and set the absolute structure of the $Pna2_1$ structure. Non-hydrogen atoms were refined anisotropically. A riding model was used to refine hydrogen atoms bound to carbon. The positions of hydrogens bound to nitrogen were taken from the difference Fourier synthesis and refined using distance restraints. The refined structures were deposited in the CSD under deposition numbers 2533931 and 2534497, respectively. We previously reported the structure of a partially hydrated form of noncentrosymmetric ZPAN (CSD ID: 2128455) [6]. A comparison of our three ZPAN structures with the original ZPAN structure (CSD ID: 795577) is provided in the supporting information in Table S1.

The crystal faces of the large ZPAN crystals were determined separately. During THz generation measurements, a permanent marker was used to label the main face of a ZPAN crystal and indicate the direction of THz polarization when maximum signal was observed. A small strip (3 mm × 0.6 mm × 0.28 mm) was carefully cut out of the marked ZPAN crystal and mounted on a polyimide loop on the goniometer with the THz polarization direction parallel to the phi axis. Enough diffraction data was collected to calculate the unit cell and orientation matrix. A crystal video was also recorded and used with the face-indexing tool in APEX6 to identify the visible faces of the crystal strip, which correspond directly to those of the parent ZPAN crystal.

**THz Generation Measurements**

We tested the THz generating abilities of ZPAN by measuring the THz generation spectrum and the refractive index from 1250 nm to 1550 nm using a Ti:Sapphire laser system with an optical parametric amplifier. THz measurements were performed using a 1-kHz repetition-rate, 800-nm Ti:Sapphire regenerative amplifier that is sent to an optical parametric amplifier that produces ~100 fs pulses across a broad range of NIR pump wavelengths from 1250 to 2150 nm. The pulses were directed through an optical chopper, which halves the repetition rate from 1 kHz to 500 Hz, and then into the THz generating crystals. Crystals were oriented repeatedly until the strongest THz signal was observed, indicating that the THz generation axis was aligned with the vertically polarized pump light. We used a 2-mm thick HDPE filter after the crystal to remove the remaining NIR and visible light, while transmitting the generated THz. We then directed the THz into a 1:5 telescope that consists of 1 in. diameter, 1 in. effective focal length gold-plated off-axis parabolic mirror (OAPM) followed by a 3 in. diameter, 5 in. focal length OAPM. The THz was then focused using a final 3 in. diameter, 2 in. focal length OAPM down to a beam with a radius of 260 μm. At the focus, we measured the electric field using electro-optic sampling with 100 fs 800 nm probe pulses and a 100 μm (110) electro-optic GaP crystal bonded to a 1 mm (001) GaP crystal.

We recorded THz generation from ZPAN crystals of varying thickness (170-730 $\mu$m). We also performed several THz transmission measurements using a 149 $\mu$m thick crystal to extract the complex THz refractive index and absorption coefficients. THz generation was recorded with ZPAN crystals at various rotation angles to evaluate the THz generation from azimuthal angles. These measurements were compared with predicted values based on calculated second order nonlinear optical coefficients (d tensor).

## RESULTS AND DISCUSSION

### Crystal Structure and Crystal Face Characterization

The structures of the centrosymmetric and noncentrosymmetric ZPAN crystals, including both the hydrated and anhydrous forms, are shown in **Fig. 2**. All structures exhibit an X-type alignment of individual molecules. However, the overlaid hyperpolarizability vectors (β) highlight the importance of noncentrosymmetric packing for THz generation; in the centrosymmetric structure (**Fig. 2(c)**), oppositely oriented β vectors will lead to destructive interference of the generated THz signal whereas the largely unidirectional alignment in the noncentrosymmetric structures **(Fig. 2(a)** and **2(b)**) will enable constructive interference.

ZPAN packing seems to be highly sensitive to crystallization conditions, and we have found that slow evaporation from acetone solutions more consistently yields ZPAN crystals with noncentrosymmetric packing. Counterintuitively, recrystallization of ZPAN from acetone containing trace amounts of water resulted in partially hydrated crystals whereas recrystallization from a mixture of acetone with 5 vol% water produced crystals without any incorporated water molecules. Despite differences in hydration, the two noncentrosymmetric crystals were visually indistinguishable and fit the same $Pna2_1$ space group as the previously reported structure (CSD ID: 795577) [13], with only slight changes in unit cell parameters (see **Table S1**). To assess the impact of water on crystal packing, we calculated the crystallographic order parameter for both structures. For the anhydrous ZPAN crystals obtained through our optimized protocol reported here, the angle ($\theta_p$) between the crystal polar axis and the molecular hyperpolarizability vector (β) is 34° (**Fig. 2(a)**), which indicates that the crystal order parameter ($\cos^3(\theta_p)$) is 0.55. The crystal order parameter and the associated angle of our partially hydrated ZPAN reported previously were identical to the anhydrous ZPAN as shown in **Fig.2(b)**. Thus, the partially incorporated water molecules do not significantly impact ZPAN packing.

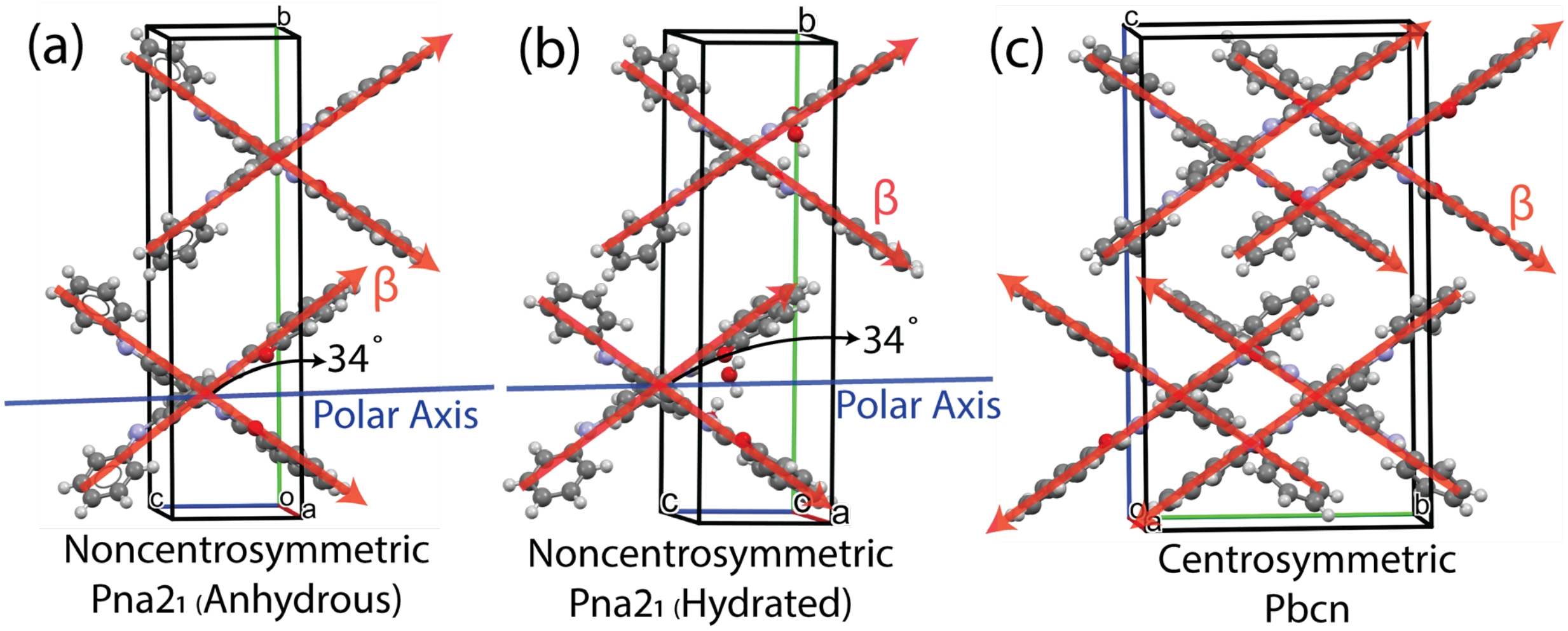

Figure 2: Molecular alignments of noncentrosymmetric anhydrous (a), hydrated (b), and centrosymmetric (c) ZPAN crystals with the hyperpolarizability vector (β) indicated by red arrows. In centrosymmetric ZPAN, the β vectors completely cancel due to the inversion symmetry whereas the β vectors in noncentrosymmetric ZPAN have a nonzero vector sum and form a 34° angle with the crystal polar axis (blue).

In addition to the molecular alignment within the crystal, the relative orientation between the polarization of the pump light and the crystal packing also has crucial impacts on THz generation efficiency. To evaluate this relative orientation in our ZPAN crystals, we indexed the main crystal faces of our anhydrous material. As **Fig. 3(a)** shows, the main crystal face is (010), and the elongated crystal growth axis is along the [001] direction. As **Fig. 3(b)** shows, the sum of the β vectors is also along the [001] direction, thus the polar axis is along the [001] direction. Accordingly, the strongest THz field should be produced when the pump IR light source is polarized along the [001] direction of the crystal, which is what was observed experimentally.

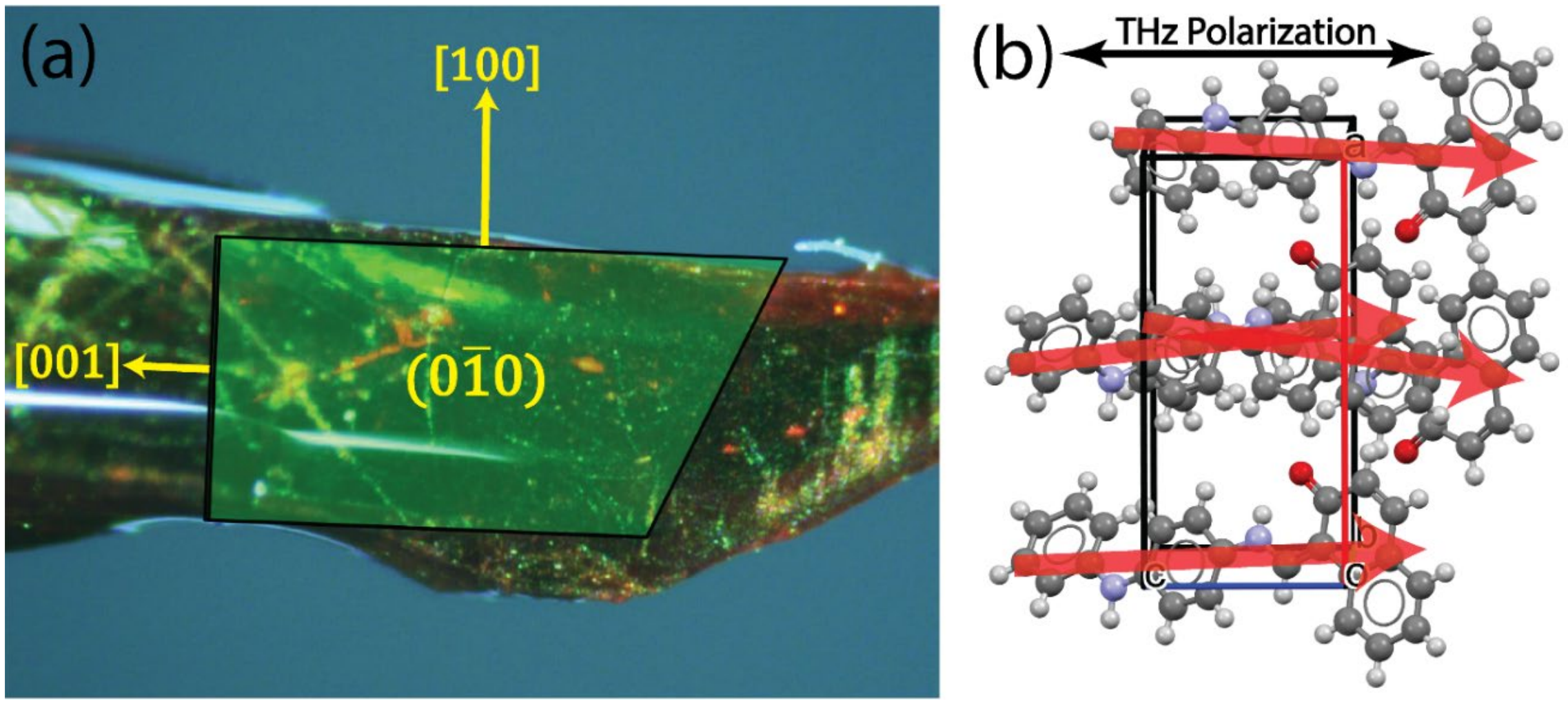

Figure 3: (a) Image from the crystal video of the ZPAN crystal strip with the main face $(0\bar{1}0)$ and perpendicular directions indicated in yellow. (b) A ZPAN unit cell viewing the main crystal growth face $(0\bar{1}0)$ oriented as outlined in part (a). The β vectors are indicated in red. The incident pump polarization and the generated THz polarization are along the black arrow.

**THz Generation Measurements**

**Fig. 4** shows the THz generated by a ~670 μm thick ZPAN crystal pumped with 1450 nm light with a 3.8 mm $1/e^2$ spot size and 0.644 mJ pulse energy, giving a fluence of 1.42 mJ/cm$^2$. The THz peak-to-peak electric field strength reaches 1 MV/cm, and ZPAN produces a strong and broad spectrum, with no major absorptions up to 3.4 THz, as shown in the inset.

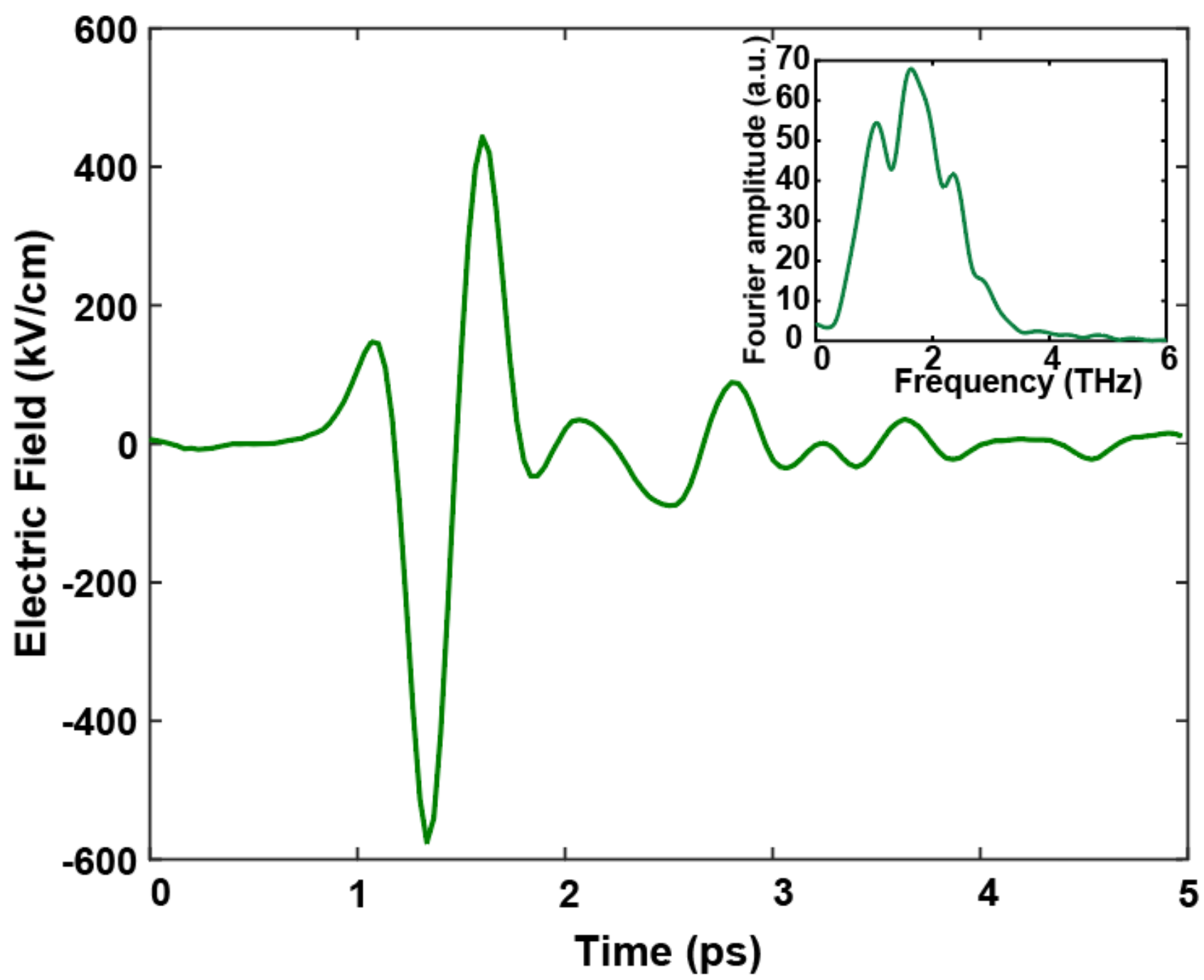

**Figure 4**: The generated THz electric field for ZPAN, with polar [001] axis aligned with the vertical IR pump polarization, measured with electro-optic sampling. Inset: The Fourier amplitude of the time-domain THz waveforms.

Wavelength and thickness dependent measurements provide a more comprehensive view of the THz generation characteristics of ZPAN. In **Fig**. **5**(**a**), we compare the THz generation spectra with wavelengths ranging from 1250-1550 nm, using 0.716 mJ pulse energy (fluence of 1.62 mJ/cm$^2$) to determine the optimal pump wavelength for ZPAN. Wavelengths 1350 to 1550 nm are all similar in spectral amplitude and bandwidth, indicating similar phase-matching. **Fig. 5(b)** shows THz generation of crystals with thicknesses ranging from 170-730 $\mu$m. As the thickness of the crystal increases, the THz spectral amplitude increases in the main range from 0-3 THz, and we observe a small decrease at higher frequencies. A strong absorption in the crystal at 3.4 THz leads to poor phase-matching at higher frequencies (see **Fig. 6** below), so there is no significant amplitude above 4 THz for any wavelength or thickness [7, 10].

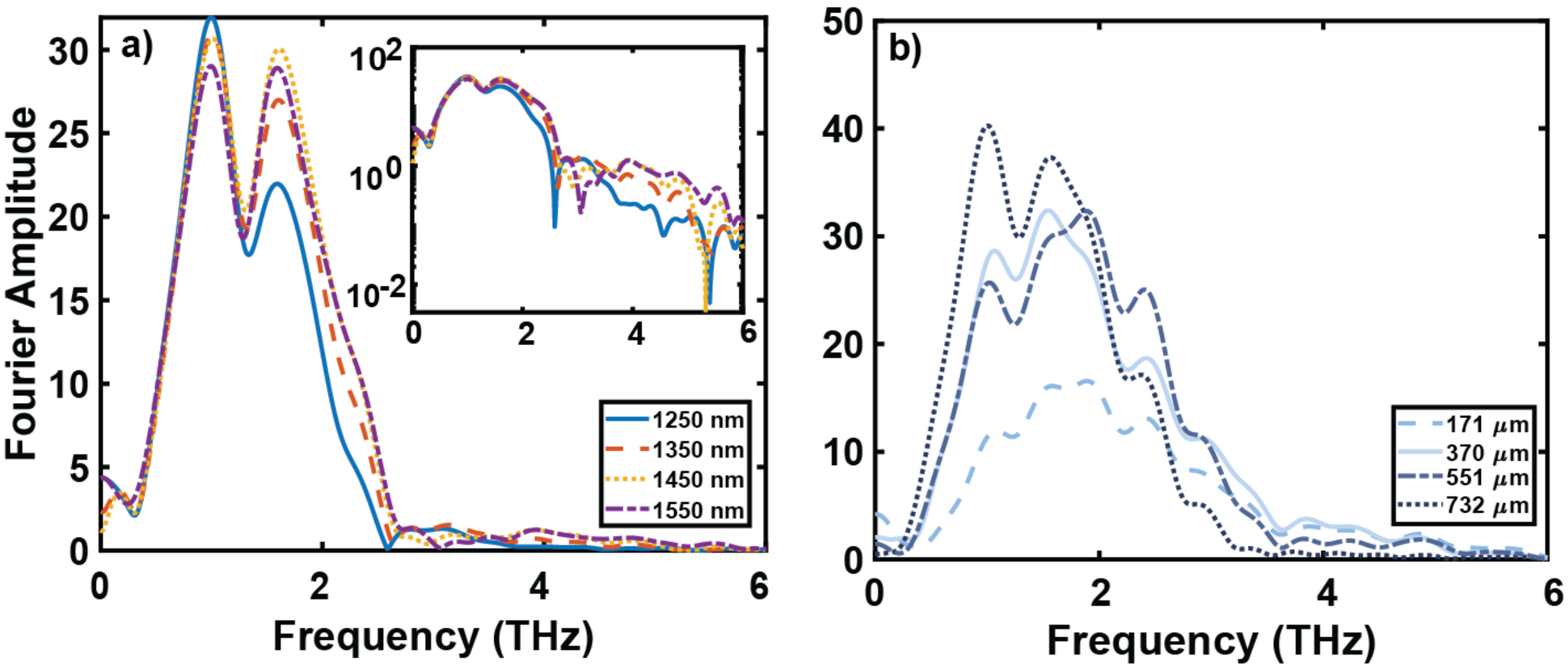

Figure 5: a) THz spectra of ZPAN pumped at different wavelengths, ranging from 1250-1550 nm. b) THz spectra generated from crystals with different thicknesses ranging from 171-732 μm. Insets shows a log vertical scale of THz amplitude.

**Refractive Index & Absorption coefficient**

We performed several THz transmission measurements to extract the complex refractive index. **Fig. 6** shows the refractive index and absorption coefficient extracted from a crystal with a thickness of 149 $\mu$m. The blue lines in **Fig. 6(a)** and **6(b)** show the refractive and absorption coefficient respectively. The dashed blue lines in **Fig. 6(a)** and **6(b)** indicate an extrapolation to 8 THz, which was not in the range of our fit. ZPAN has a low-frequency THz refractive index of approximately 2.2 with a main absorption feature at ~3.4 THz. The absorption at 3.4 THz, corresponding to a larger step in the refractive index, is strong enough that little to no THz is generated above this frequency due to the poor phase-matching. The yellow dashed line in **Fig. 6(b)** shows the infrared absorption (see upper wavelength axis). We also determined the group index of the crystal at various wavelengths via autocorrelator measurements [11]. The data is represented by the red dots in **Fig. 6(a)**, which show the average values and the associated error. To have good phase-matching and efficient THz generation, the group velocity of the IR light must match the phase velocity of the THz wave propagating through the crystal medium. We observed in **Fig. 5(a)** that the pump wavelengths ranging from 1350-1550 nm are similar in spectral amplitude and width and are well phase matched below 3 THz. Comparing this observation to the data in **Fig. 6(a)**, we see that within the error bars, all the group indices are well matched to the THz refractive index of the material below 3 THz. The consistent phase matching at lower frequencies gives ZPAN the advantage of producing a consistent output of THz across the tested range of wavelengths.

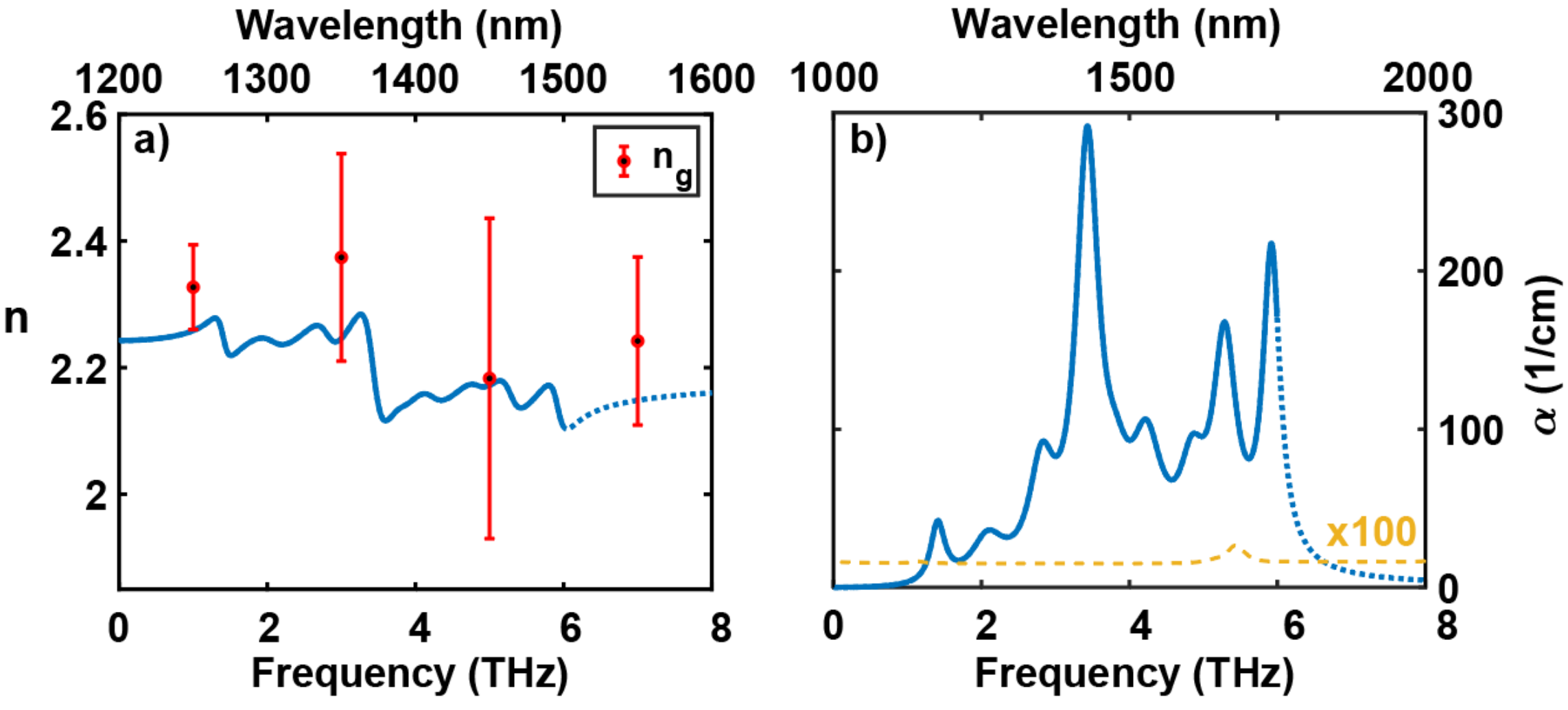

Figure 6. a) The extracted refractive index (blue line) and the group index (red circles) extracted from auto correlator measurements. b) The extracted THz absorption coefficient (blue line) and IR pump absorbance (yellow dashed line) as a function of wavelength. The dashed line from 6 THz to 8 THz indicates an extrapolation outside the fit range of the model.

## Comparison with Other Common NLO Crystals

In our previous work [6], we compared preliminary THz generation from ZPAN with DAST, OH1 and PNPA crystals. In subsequent work [7], we showed that with higher quality PNPA crystals, we can generate higher field strengths than was reported in Ref [6]. We found the same trend here; higher quality ZPAN crystals show a 30-40% increase in peak field strength compared to our earlier results. For comparison purposes, **Fig. 7** shows the THz waveforms and spectra for several common THz generators, including BNA, DAST, GaP and ZPAN. Each crystal was pumped at 1450 nm for comparison. ZPAN has a peak-to-peak electric field strength that is near 1 MV/cm, which is comparable to that of DAST and higher than that of BNA [7, 10].

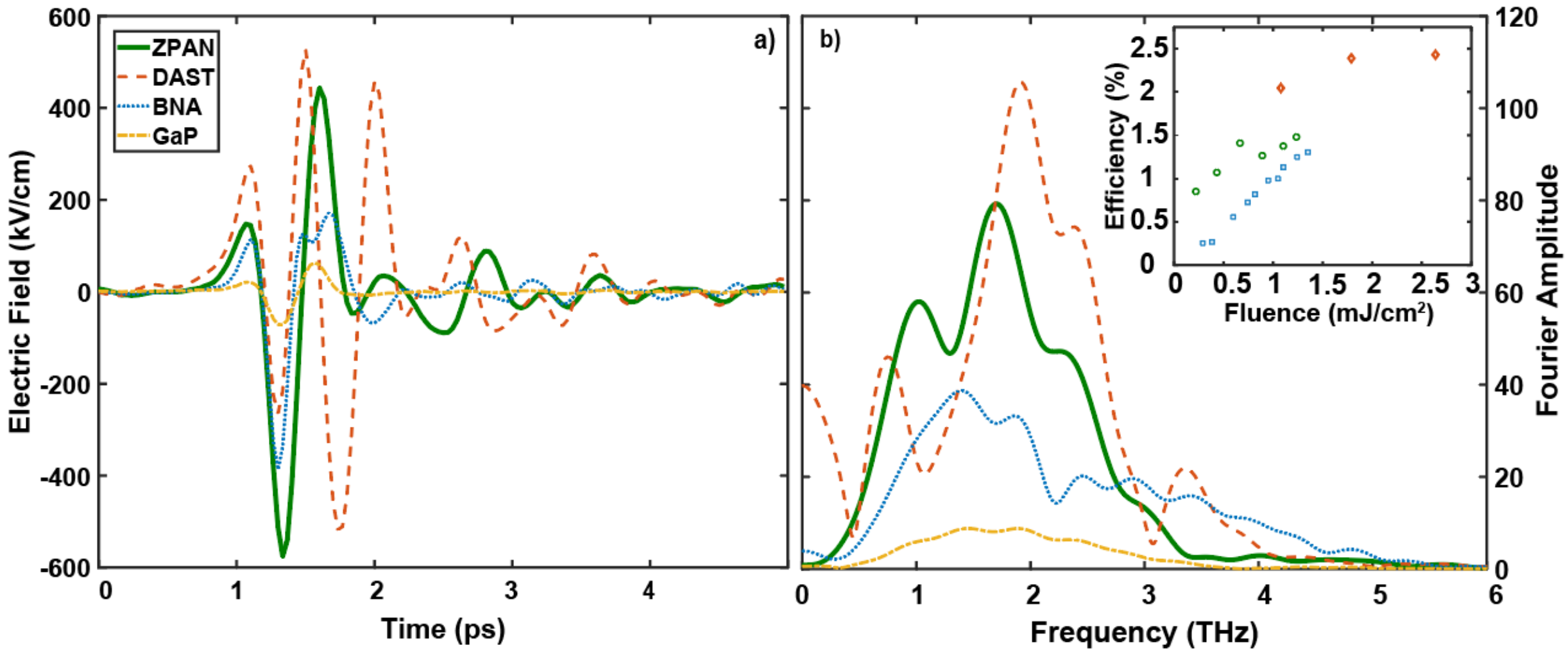

Figure 7: a comparison of ZPAN with other commercially available crystals, BNA, DAST and GaP at 1450 nm. a) The generated electric field for each crystal. b) THz spectra for each of the tested crystals. Inset: The generation efficiency as a function of laser fluence for BNA (blue squares), ZPAN, (green circles) and DAST (red diamonds). BNA and DAST efficiencies come from Ref. [9] and Ref. [10] respectively.

One clear advantage that ZPAN offers is the ability to grow much larger crystals than other organic materials like DAST, DSTMS, OH1, and PNPA. Typically, the largest crystals that can be grown with these other materials is an 8-10 mm aperture, and even those apertures are extremely difficult to produce. With high-power laser systems, small apertures limit the pump pulse energy that can irradiate the THz generation crystal without damage. Larger ZPAN crystals circumvent this limitation by utilizing the full output power of such an intense laser system. In **Fig. 8** we show a simple comparison of the theoretical THz output of DAST and ZPAN as a function of crystal aperture. The theoretical THz output is normalized for a DAST crystal with a maximum aperture of 10 mm diameter, assuming pumping with the fluence that gives maximum efficiency (~2 mJ/cm$^2$, see **Fig. 7**). As **Fig. 8** shows, DAST with its ~2.4% generation efficiency will generate more THz than ZPAN (~1.7% generation efficiency) under the same pumping conditions and crystal size. However, because larger ZPAN crystals can be produced, total THz output from ZPAN can be increased beyond that of DAST by scaling up laser power and crystal aperture simultaneously. For example, a 12-mm ZPAN aperture could provide the same THz output as a 10-mm DAST aperture with output increasing quadratically as a function of aperture size, all while keeping the pump fluence the same and avoiding damage. This is not practical for DAST or other common THz crystals due to their limited crystal size. As a result, ZPAN can be used with higher power laser systems to generate significantly stronger THz pulses than any other organic crystals currently being produced.

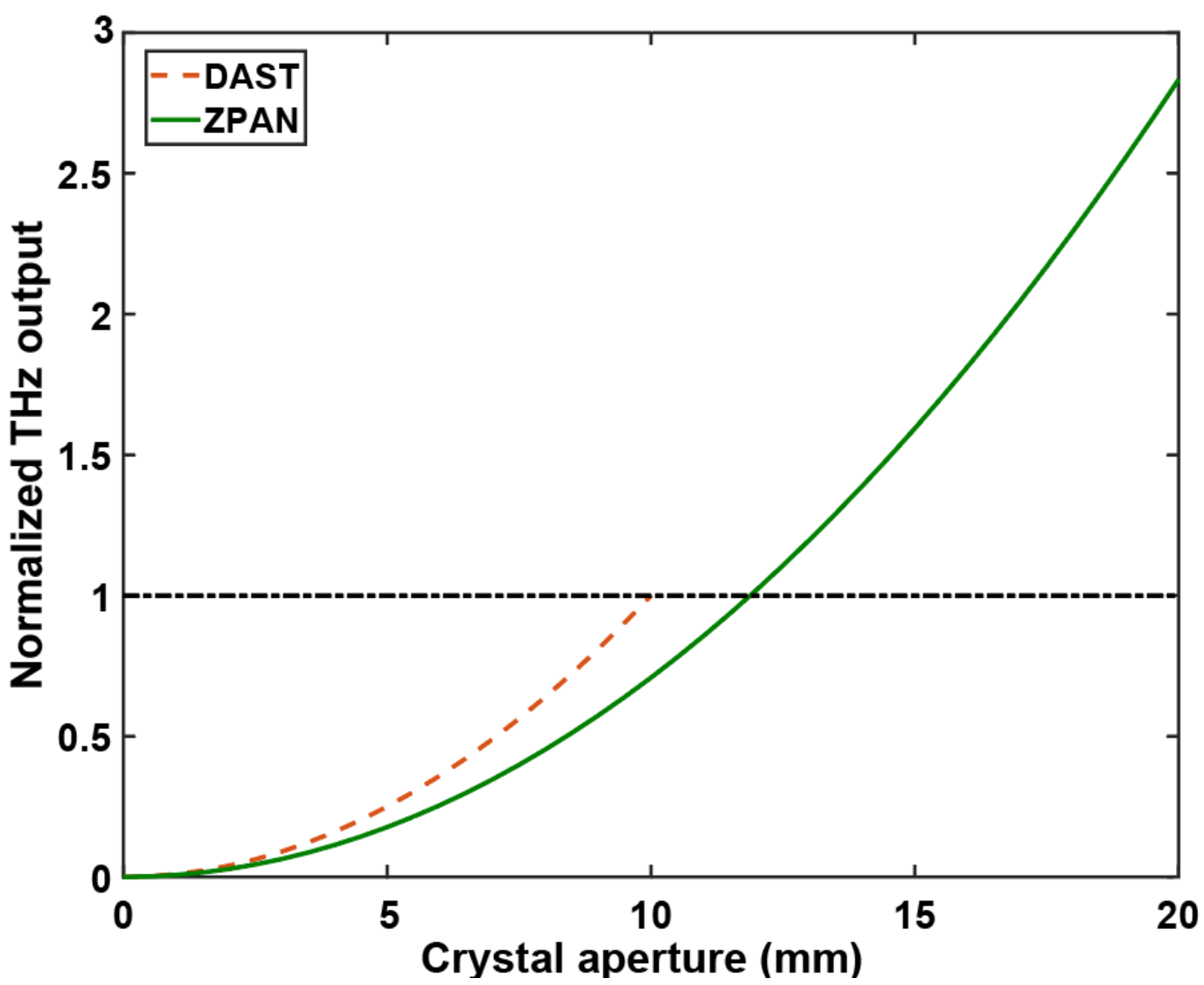

Figure 8. A comparison of THz output (pulse energy) between DAST and ZPAN at a fixed fluence for maximum generation efficiency (~2 mJ/cm$^2$). The y-axis is normalized to the THz output from a DAST crystal with a maximum diameter of 10 mm.

## Theoretical and Experimental THz Generation Capabilities from Different Crystal Faces

In the previous sections, we examined the crystal structures, optical properties, and THz generation of ZPAN, demonstrating that it is a viable NLO crystal for THz generation. To further understand the THz generation of ZPAN, we now explore the theoretical aspects of its NLO properties, examining the influence of pump polarization and crystal orientation on the THz emission. Because the noncentrosymmetric ZPAN crystal structure has no inversion center, the sum of its molecular vectors varies in different directions. Thus, the nonlinear optical properties induced from different crystal faces can vary widely. To extend our understanding of the THz generation capabilities from different crystal faces of ZPAN, we calculated the second order nonlinear optical coefficients ($d_{IJK}$) under Kleinman condition for optical rectification and crystal electric polarization based on the equations below [17]:

$$[d_{IJK}(0\omega;\omega,\omega)] = [N f_I^{0\omega} f_J^{\omega} f_K^{\omega} \sum_{s=1;i,j,k}^{n} \cos(I,i(s))\cos(J,j(s))\cos(K,k(s))\,\beta_{ijk}(s)] \quad (1)$$

$$\begin{bmatrix} P_x \\ P_y \\ P_z \end{bmatrix} = 2\epsilon_0 [d_{IJK}(0\omega;\omega,\omega)] \begin{bmatrix} E_x^2 \\ E_y^2 \\ E_z^2 \\ 2E_yE_z \\ 2E_xE_z \\ 2E_xE_y \end{bmatrix} \quad (2)$$

In Eq. 1, the indices *I, J, K* represent the Cartesian axes of crystallographic frame, and *i, j, k* represent the molecular axes. *N* is the number of unit cells per unit volume. *n* is the number of molecules inside a unit cell and $\beta_{ijk}(s)$ are the molecular hyperpolarizability tensor elements. The summation of the product of the cosine terms accounts for the conversion between the molecular reference frame and the crystal reference frame for all *n* molecules. $f_I^{0\omega}$, $f_J^{\omega}$ and $f_K^{\omega}$ are Lorentz local-field factors, which are related to the refractive index by $f_I^{\omega} = \frac{1}{3}[(n_I^{\omega})^2 + 2]$. To accurately calculate these factors, the refractive indexes of all three main crystal faces are required. However, it is not feasible to grow crystals in three directions with the same quality, as crystals often have a preferred growth direction [18]. Since the refractive index along the THz generation axis of ZPAN is very similar to that of the NLO crystal DAST, we approximated the refractive indices of the other two principle optical axes using the value for DAST, which is 1.6 [19, 20]. In Eq. 2, the $P_i$ terms represent polarizations contributed from individual crystallographic directions and the $E_i$ terms account for the polarization of input light. The calculated nonlinear optical coefficients in pm/V are shown below.

$$\begin{pmatrix} 0 & 0 & 0 & 0 & 4.62 & 0 \\ 0 & 0 & 0 & 57.44 & 0 & 0 \\ 4.62 & 57.44 & 117.64 & 0 & 0 & 0 \end{pmatrix}$$

The largest coefficient is $d_{333}$=117.64 pm/V. For comparison, inorganic NLO crystal – $LiNbO_3$ has its largest coefficient around 25.6 pm/V [21] and DAST has its largest coefficient ranging from 210 pm/V to 1010 pm/V [22]. While the nonlinear optical coefficient of ZPAN is not larger than DAST, as confirmed by THz generation efficiencies reported above, the value is larger than $LiNbO_3$.

Applying the calculated nonlinear coefficients to Eq. 2, the polarization with respect to x, y, z crystallographic directions ($P_x$, $P_y$, $P_z$) were calculated as shown in table 8. The total polarization of the three main crystal faces of ZPAN (denoted as $P_{100}$, $P_{010}$ and $P_{001}$) were calculated by taking the vector sum of the relevant polarization components $P_x$, $P_y$ and $P_z$. For instance, the total polarization of the (100) face was determined by $P_{100}$=$P_y^2$+$P_z^2$; and similarly for the other faces. The electric field strength used in the calculation is 2.88 MV/cm, which was calculated based on our Ti:sapphire laser (0.7mJ pulse energy and 100fs pulse duration).

Table 1: Calculated polarization values with respect to the three main crystal faces.

| Polarization | $mC/m^2$ |
|---|---|
| $P_{100}$ | 17.2 |
| $P_{010}$ | 17.2 |
| $P_{001}$ | 0 |

As **Table 1** shows, the $P_{001}$ value is zero, which indicates little to no nonlinear effect can occur from the (001) face. This is consistent with our expectation based on the crystal packing. As shown in **Fig. 9(c)**, when only the projection of molecules onto the (001) crystal face is considered, we can observe that the molecules viewed in this direction are oppositely oriented, thus the hyperpolarizability vectors cancel each other, and the vector sum is zero for the (001) face.

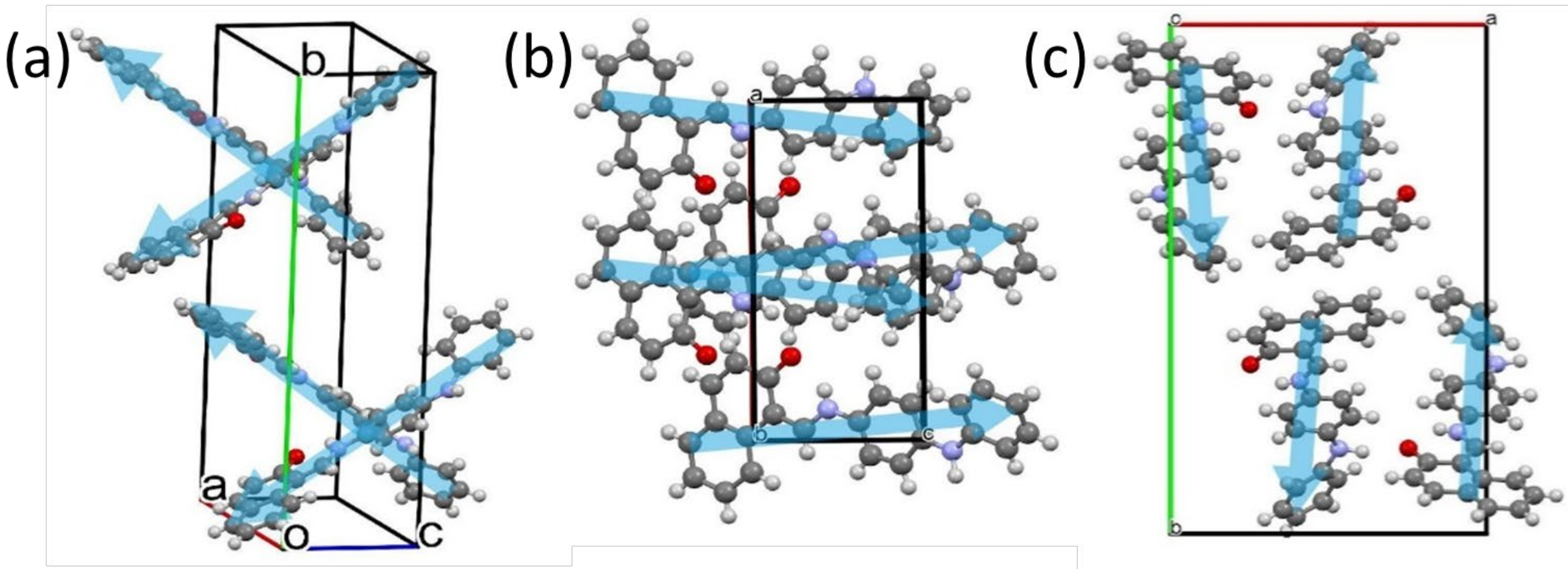

Figure 9: Molecular alignment viewed from the (100) (a), (010) (b) and (001) (c) crystal faces. The vectors (blue) indicate the hyperpolarizability vectors. The sums of the hyperpolarizability vectors are non-zero for both 100 and 010 face, while the sums of the vectors are zero for the 001 face.

In contrast, the calculated $P_{100}$ and $P_{010}$ values are nonzero and equal because the sum of the hyperpolarizability vectors is nonzero as illustrated in **Fig. 9 (a-b)**, indicating that THz generation performance for those two faces should theoretically be the same. We note that the Lorentz local-field factors were treated the same for both crystal faces, so it is possible that the two different faces would have slightly different THz output if a ZPAN crystal with (100) as the main face could be grown to obtain more accurate information for that face. In addition, even though the two faces have the same polarization values, the off-diagonal tensor components are very different, where $d_{322}$ = 57.4pm/V for the (100) face and $d_{311}$ = 4.6 pm/V for the (010) face. This suggests that if the (100) crystal face could be grown to sufficient size, ZPAN could be used for THz generation from the off-diagonal nonlinear component [23].

**Crystal Face testing results**

To confirm the theoretical d-tensor calculations, we altered the polarization of the pump light that irradiates the crystal over a range of 0-200 degrees. The GaP EO-detection crystal is set to detect vertically polarized light, so we measure the vertically polarized components of the THz field. The addition of a λ/2 waveplate mounted before the THz generation crystal enables us to alter the incoming polarization of the pump, which modifies the specific nonlinear coefficients that contribute to THz generation, as described in Eq. 2. ZPAN crystals exhibit (010) as the main face, so we can select the nonlinear coefficients from the z- and x- directions, or [001] and [100] respectively, depending on which orientation of the crystal is vertical. Combining the calculated nonlinear coefficients with various polarization directions of the incoming light, we modeled which components of polarization contributed to the generation of THz as shown in **Fig. 10**. When the THz generation crystal is on its [001] generation axis, the model and experimental data fit well. When we rotate the crystal 90 degrees so that the [100] direction is vertical, the field strength is minimal as expected, though the model deviates slightly from the experimental data potentially due to incomplete attenuation of the non-vertical components from the THz polarizers that are present after the generation crystal.

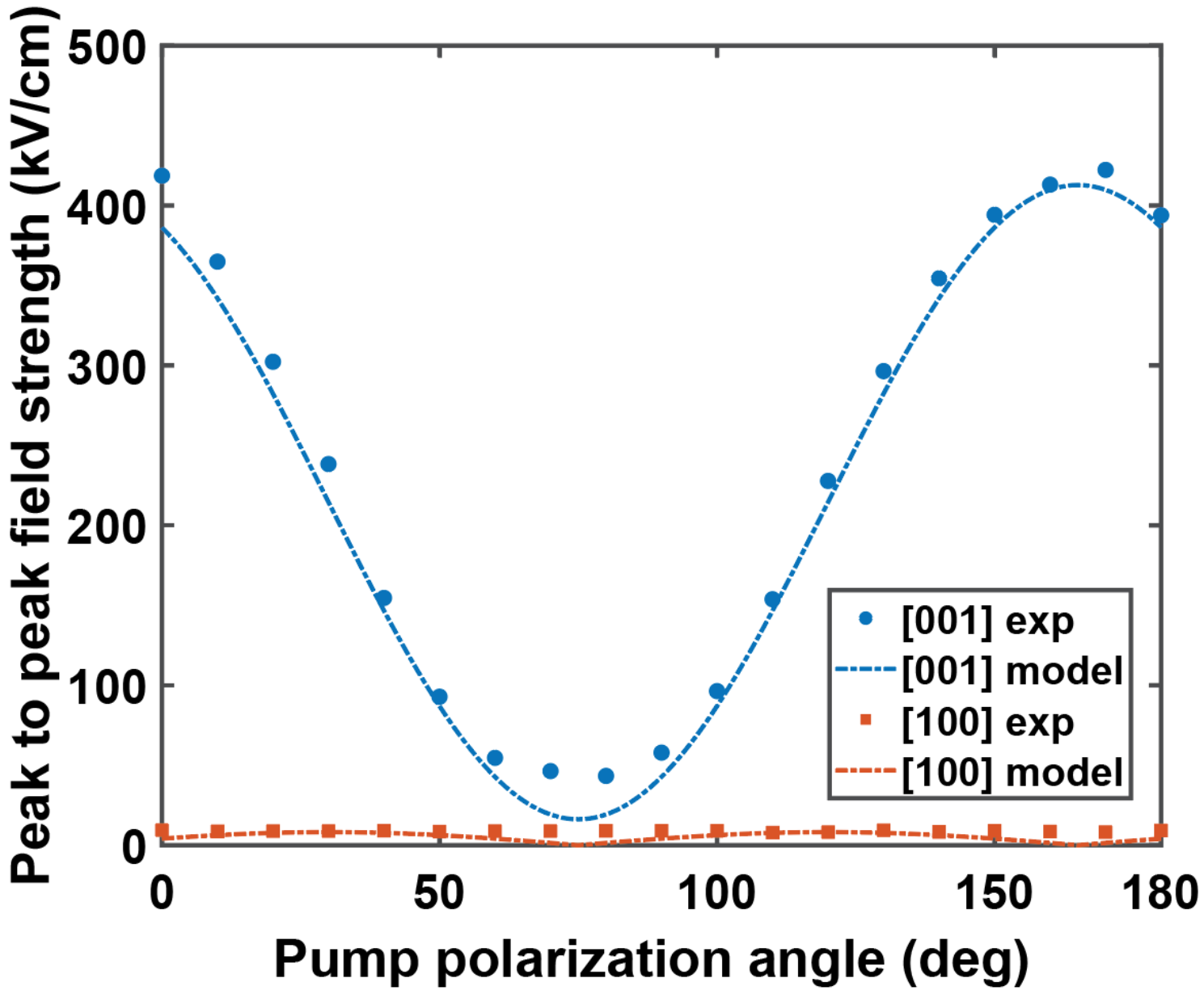

Figure 10. Measured (dots) and modeled (lines) peak-to-peak electric field strengths generated by a ZPAN crystal as the pump polarization angle rotates between 0° and 180° alignment with either the [001] direction (in blue) or the [100] direction (in green).

## CONCLUSION

In summary, we report an optimized synthesis and crystallization strategy for ZPAN that enables reproducible growth of large, high quality single crystals capable of high field terahertz generation. Noncentrosymmetric packing in $Pna2_1$ was most reliably achieved by recrystallizing from a mixed acetone:water solvent system (5 vol% water), yielding rectangular crystals as large as 7 cm × 1 cm × 0.2 cm. Structural analysis showed that the main crystal face is (010) with the elongated crystal growth axis and polar axis aligned along the [001] direction. Consistent with this symmetry, the strongest THz emission was observed when the incident IR pump polarization direction was parallel to [001].

The developed ZPAN crystals can generate an electric field of near 1 MV/cm with a smooth spectrum from 0.5-3.4 THz. Wavelength- and thickness-dependent measurements show that ZPAN has a consistent THz generation across pump wavelengths 1350 to 1550 nm due to similar group velocities. Refractive index and absorption coefficient measurements show a strong absorption at ~3.4 THz that defines the upper limit of THz generation. The calculated second order nonlinear optical coefficients and predicted crystal polarizations for the three main crystal faces indicate that both (100) and (010) are theoretically capable of THz generation when the pump polarization is along the [001] direction, consistent with the experimental generation results.

A key development in this work is the ability to grow much larger crystals of ZPAN than other organic materials like DAST, DSTMS, OH1, and PNPA; apertures larger than a centimeter in scale enable the total THz output from ZPAN to be increased beyond that of any of these other materials by jointly scaling laser power and crystal size. This capability makes ZPAN a compelling candidate for use with very high-pulse energy laser systems in the 1-3 THz range.

**Supplementary Material Summary**

S1: Comparison of X-ray crystal structures of ZPAN

**Acknowledgements**

The authors acknowledge the National Science Foundation (DMR 2104317) and (PFI 2345726) for the support of this research.

**Conflict of Interest**

JAJ and DJM are founders of Terahertz Innovations, LLC, which sells THz generation materials commercially.

**Author Contributions**

**M.J. Lutz:** Data curation (equal); Formal Analysis (equal); Investigation (equal); Methodology (equal); Validation (equal); Visualization (equal); Writing (equal); **S.H. Ho:** Data curation (equal); Formal Analysis (equal); Investigation (equal); Methodology (equal); Validation (equal); Visualization (equal); Writing (equal); **Z.B. Zaccardi:** Data curation (supporting); Formal Analysis (supporting); Investigation (supporting); Methodology (supporting); Validation (supporting); Visualization (supporting); Writing (supporting); **P. Peterson:** Data curation (supporting); Formal Analysis (supporting); Investigation (supporting); Methodology (supporting); Validation (supporting); Visualization (supporting); **E. Jones:** Data curation (supporting); Formal Analysis (supporting); Investigation (supporting); Methodology (supporting); Validation (supporting); Visualization (supporting); **S.J. Smith:** Data curation (supporting); Formal Analysis (supporting); Investigation (supporting); Methodology (supporting); Validation (supporting); Visualization (supporting); Writing (supporting); **D.J.**

**Michaelis:** Conceptualization (equal); Data curation (equal); Formal Analysis (equal); Investigation (equal); Methodology (equal); Project administration (equal); Resources (equal); Supervision (equal); Validation (equal); Visualization (equal); Writing (equal); **J.A. Johnson:** Conceptualization (equal); Data curation (equal); Formal Analysis (equal); Funding acquisition (equal); Project administration (equal); Resources (equal); Supervision (equal); Methodology (equal); Writing (equal);

## Data Availability Statement

The data that support the findings of this study are available from the corresponding author upon reasonable request.

**Supporting Information for**

# ZPAN: a New Organic NLO Crystal for High Intensity THz Generation

Matthew J. Lutz, (Enoch) Sin-Hang Ho, Zachary B. Zaccardi, Paige Petersen, Elisha Jones, Stacey J. Smith, David J. Michaelis*, Jeremy A. Johnson*

Department of Chemistry and Biochemistry, Brigham Young University, Provo, UT 84602, USA

* Electronic Mail: dmichaelis@chem.byu.edu, jjohnson@chem.byu.edu

# Table of Contents

Page #

## Contents

## 1. Comparison of X-ray crystal structures

**Table S1.** Comparison of X-ray crystallographic information for the originally reported ZPAN structure (CCDC 795577) and our three ZPAN structures (anhydrous, partially hydrated, and centrosymmetric).

| **Compound** | Original ZPAN [13] CCDC 795577 | ZPAN (anhydrous) CCDC 2533931 | ZPAN [6] (partially hydrated) CCDC 2128455 | ZPAN (centrosymmetric) CCDC 2534497 |
|---|---|---|---|---|
| Formula | $C_{23}H_{18}N_2O$ | $C_{23}H_{18}N_2O$ | $C_{23}H_{18.15}N_2O_{1.07}$ | $C_{23}H_{18}N_2O$ |
| Formula weight | 338.39 | 338.39 | 339.73 | 338.39 |
| Temperature | 296(2) K | 100(2) K | 100(2) K | 100(2) K |
| Theta range | 1.56 to 27.08 (Mo X-rays) | 6.928 to 74.437° (Cu X-rays) | 3.953 to 74.461 (Cu X-rays) | 4.008 to 82.377° (Cu X-rays) |
| Crystal system | Orthorhombic | Orthorhombic | Orthorhombic | Orthorhombic |
| Space Group | Pna21 | Pna21 | Pna21 | Pbcn |
| Point group | mm2 | mm2 | mm2 | mmm |
| a (Å) | 13.0688(8) | 12.9793(4) | 12.9872(4) | 12.1847(4) |
| b (Å) | 22.5362(13) | 22.0298(6) | 21.9914(6) | 12.7718(4) |
| c (Å) | 6.0975(3) | 6.1114(2) | 6.1207(2) | 22.0608(7) |
| α = β = γ (˚) | 90 | 90 | 90 | 90 |
| V (Å$^3$) | 1795.84(17) | 1747.44(9) | 1748.11(9) | 3433.11(19) |
| Z | 4 | 4 | 4 | 8 |
| Density (g/cm$^3$) | 1.252 | 1.286 | 1.286 | 1.309 |
| # of total reflections | 12478 | 27617 | 28930 | 39394 |
| # of unique reflections | 1940 | 3347 | 3434 | 3381 |
| R | 0.0554 | 0.0340 | 0.0293 | 0.0593 |
| wR2 | 0.0650 | 0.0914 | 0.0718 | 0.1188 |
| Flack (Parsons' method) | - | -0.051(178) | 0.020(113) | - |
| Flack (Classical Fit) | - | 0.208(308) | 0.018(318) | - |

- A Flack parameter cannot be calculated for the centrosymmetric structure, and [13] did report a Flack parameter.